\documentclass[a4paper,fleqn,usenatbib]{mnras}

\usepackage{newtxtext,newtxmath}

\usepackage[T1]{fontenc}
\usepackage{ae,aecompl}

\usepackage{graphicx}	
\usepackage{amsmath}	
\usepackage{amssymb}	
\usepackage[normalem]{ulem}
\useunder{\uline}{\ul}{}
\usepackage{booktabs}

\usepackage{etoolbox}
\makeatletter
\patchcmd\@combinedblfloats{\box\@outputbox}{\unvbox\@outputbox}{}{%
   \errmessage{\noexpand\@combinedblfloats could not be patched}%
}%
 \makeatother

\title[SMBH accretion in starburst galaxies.]{Revealing the differences in the SMBH accretion rate distributions of starburst and non-starburst galaxies.}
\author[Grimmett et al.]{
L.P. Grimmett$^{1}$,\thanks{E-mail: lpgrimmett1@sheffield.ac.uk}
J.R. Mullaney$^{1}$,
S. Jin$^{2,3}$,
E. Bernhard$^{1}$,
E. Daddi,$^{2}$
K. Walters${^4}$
\\
$^{1}$Department of Physics and Astronomy, University of Sheffield, Sheffield S3 7RH, UK\\
$^{2}$CEA, IRFU, DAp, AIM, Universit\'e Paris-Saclay, Universit\'e Paris Diderot, Sorbonne Paris Cit\'e, CNRS, F-91191 Gif-sur-Yvette, France\\
$^{3}$ School of Astronomy and Space Science, Nanjing University, Nanjing 210093, People\textquotesingle s Republic of China\\
$^{4}$School of Mathematics and Statistics, University of Sheffield, Sheffield S3 7RH, UK
}

\date{Accepted XXX. Received YYY; in original form ZZZ}

\pubyear{2018}

\begin{document}
\label{firstpage}
\pagerange{\pageref{firstpage}--\pageref{lastpage}}
\maketitle

\begin{abstract}
We infer and compare the specific X-ray luminosity distributions for a sample of massive (i.e. $\log_{10} (M*/M\odot) > 10.5$) galaxies split according to their far-infrared-derived star-forming properties (i.e., starburst and non-starburst) and redshift. We model each distribution as a power-law with an upper and lower turnover, and adopt a maximum likelihood method to include information from non-detections in the form of upper limits. When we use our inferred distributions to calculate the ratios of high to low sLx AGN (corresponding to above and below $0.1\lambda_{\text{Edd}}$, respectively) we find that starbursts have significantly higher proportions of high sLx AGN compared to their non-starburst counterparts. These findings help explain the increase in average X-ray luminosity in bins of increasing SFR reported by previous studies.
\end{abstract}




\section{Introduction}
The now widely-accepted correlation between various physical properties of galaxies and their central supermassive black holes (hereafter, SMBH) suggests some level of co-evolution between the two \citep[e.g.,][]{Magorrian98,Marconi03,Haring04,Gultekin09}. However, the precise nature of this co-evolution is not yet fully understood and is therefore the focus of much ongoing research. A common method of exploring the link between galaxies and their resident SMBHs has been to investigate how the two have grown together over time. To this end, there have been many recent studies exploring the relationship (or lack thereof) between SMBH growth (witnessed as active galactic nuclei; hereafter AGN) and the rate of galaxy growth via star formation \citep[e.g.,][]{Rosario12,Mullaney12jan,Chen13,Feltre13,Azadi15,Delvecchio15,Stanley15,Bernhard16,Pitchford16, Aird18b,Bernhard19, Yang18}.

Early statistical studies into the relationship between SMBH growth and star formation largely relied on star formation rates (hereafter, SFR) based on optical data, whether from spectroscopy (e.g., [OII], \citealt{Kauffmann03, Silverman08}) or optical colours or spectral energy distribution (SED) fits \citep[e.g.,][]{Mainieri11, Aird12}. These studies generally found that AGN preferentially (but not exclusively) reside in star-forming galaxies. However, with optical wavelengths affected by dust obscuration and potential AGN contamination, research in this area has recently benefited greatly from the availability of far-infrared (FIR) data for large numbers of AGN, particularly from \textit{Herschel} and ALMA. Being largely impervious to dust extinction and, aside from the most extreme cases, largely unaffected by AGN contamination \citep[e.g.,][]{Mullaney11, Mullaney12jan} FIR wavelengths are, in many respects, an ideal measure of SFR in AGN. Recent studies have exploited this to reaffirm AGN's preference for star-forming galaxies \citep[e.g.,][]{Mullaney12jan, Rosario13, Bernhard16}. Beyond this, other studies utilising FIR wavelengths sought to identify whether AGN luminosity (a proxy for SMBH growth) and SFR are correlated \citep[e.g.,][]{Rosario12,Chen13,Feltre13,Azadi15,Delvecchio15,Stanley15,Pitchford16}. Since not all AGN are detected at FIR wavelengths, such studies have adopted averaging techniques (in particular stacking but also survival analysis, e.g., \citealt{Stanley15}) to account for non-detections. In general, these studies have found little evidence of a strong correlation between SFRs averaged in bins of AGN luminosity.

The lack of a strong correlation between AGN luminosity and SFR may appear at odds with a co-evolution between SMBH and their host galaxies. However, \cite{Hickox14} demonstrated that this flat relationship could arise due to the variability of AGN on timescales shorter than the variation of galaxy SFR. Indeed, studies that derive the average AGN luminosity in bins of SFR or galaxy mass -- effectively averaging over the short term variability in bins of the more stable quantity -- do find a stronger positive correlation between AGN luminosity and SFR \citep[e.g.,][]{Mullaney12jul, Chen13, Azadi15, Lanzuisi17, Brown18}. However, \cite{Rodighiero15} report that while the average AGN luminosity of starburst galaxies is, indeed, higher than in ``normal'' (i.e., main-sequence) star-forming galaxies, it is not proportionately so (i.e., the ratio of AGN luminosity to SFR in starburst galaxies is {\it lower} than in main-sequence galaxies), although it is not yet clear whether this could be due to other effects such as high levels of absorption in starburst galaxies.

The finding that average AGN luminosity increases with SFR implies that the distribution of AGN luminosity changes as a function of the star-forming properties of the host galaxy. However, averages (such as linear means, medians and modes) give little insights into the full shape of these distributions. For example, does a sample have a higher average AGN luminosity because each AGN is slightly more luminous, or is it due to a small number of extreme, high luminosity AGN pulling the average up? Addressing such questions will provide a deeper understanding of the relationship between SMBH growth and galaxy growth:  is the heightened average in star-forming galaxies caused by a slight increase in the activity of all AGN or a greater fraction of extreme cases? A direct way of addressing this is to determine how the AGN luminosity distribution changes as a function of the star-forming properties of their host galaxies. This has been explored in some recent studies \citep[e.g.,][]{Aird12,Azadi15,Wang17,Aird17} who used rest frame optical to near infrared colours or SED fitting routines to identify samples of star-forming and quiescent galaxies and determined the specific AGN X-ray luminosity (i.e., X-ray luminosity per unit host stellar mass, hereafter $sL_{\rm X}$, which is commonly used as a proxy for Eddington ratio) distributions for each sample. In general, these studies report a suppression of AGN activity in quiescent galaxies, particularly at modest specific AGN luminosities (i.e., equivalent to Eddington ratios of $\lambda_{\text{Edd}}\sim0.1$). However, in light of the aforementioned difficulties associated with SFR estimates derived from optical wavelengths, it has yet to be determined whether these results are also observed when using FIR-derived SFRs.

In this study, we measure the full (i.e. including detected and undetected sources) $sL_{\rm X}$ distributions of galaxies whose star-forming properties have been measured from FIR data. We then compare these distributions of starbursting galaxies (defined by their specific SFR, i.e., SFR per unit stellar mass) against non-starbursting galaxies. In order to obtain sufficient source statistics we use the catalogue presented by \cite{Laigle16} containing sources within the COSMOS field and supplement it with FIR SFRs (see Section~\ref{data}). To measure the AGN luminosity distributions we construct a flexible model (see Section~\ref{model}) that allows for both a power law style distribution (with lower and upper exponential turnovers) and a distribution that is more log-normally shaped allowing the data to determine which is more appropriate.  Finally, we present the complete results and potential explanations in section~\ref{results} and possible implications and caveats in section~\ref{discussion}. Throughout we assume a 6-parameter $\Lambda$CDM cosmological model, with parameter values best inferred from the WMAP 9-year observations \cite{Hinshaw13}.

\section{Data}\label{data}

We start this section by summarising the process by which we derived our final sample of galaxies before elaborating on the details of this process in the subsections (i.e., stellar mass, SFR) that follow.

\subsection{Sample selection} \label{select}

To measure the AGN luminosity distributions it is important that we have as clean and unbiased a sample as possible. This is most easily obtained by using blank field surveys. In addition we also require a large sample, to avoid suffering from small sample size statistics, and comprehensive multi-wavelength coverage (for deriving stellar masses and SFRs). In particular, we also require good X-ray coverage as this provides, arguably, the most uncontaminated measure of AGN luminosity \citep{Brandt15}. These requirements are well-met by the Cosmic Evolution Survey \citep[COSMOS,][]{Scoville07} making this a natural choice for this study.

\begin{figure}
  \includegraphics[width=\columnwidth]{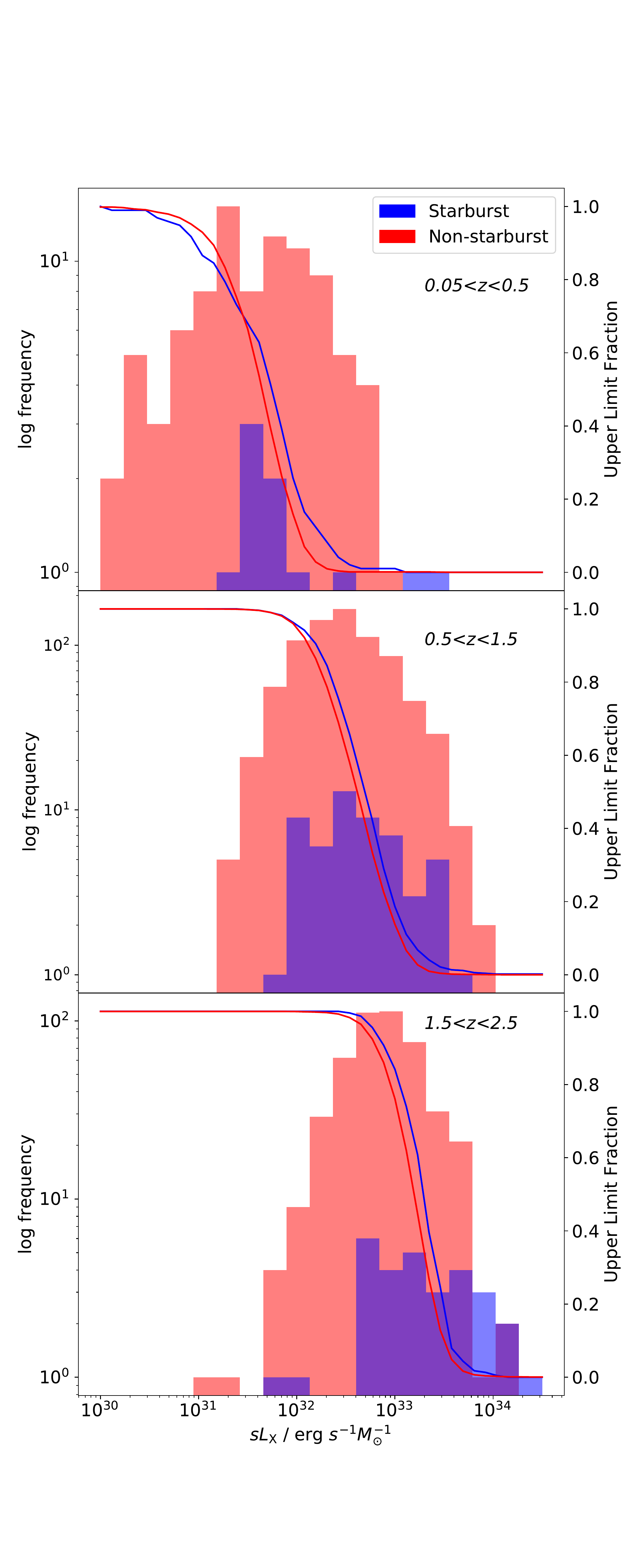}
  \caption{The detected $sL_{\rm X}$ distribution for the starburst (blue histogram) and non-starburst (red histogram) samples. Also shown is the cumulative upper limit fraction for the starburst (blue line) and non-starburst (red line). This illustrates where information about the true distribution is likely to come from (i.e., whether predominantly from the detections or non-detections).}
  \label{hist_ulfrac}
\end{figure}

Our sample selection starts with the catalogue presented by \citet[L16 from herein]{Laigle16}, which contains photometric data for 1,182,108 sources in the COSMOS field. We supplement this with X-ray data from the catalogue presented by \citet[C16 from herein]{Civano16}, which contains X-ray fluxes from \textit{Chandra} for 4016 sources. We then apply the following steps to derive our final sample:

\begin{enumerate}
  \item Firstly we ensure that the redshifts between L16 and C16 are consistent. We start with the photometric redshifts presented in L16 for all our sources as default. Then, for those sources present in C16, we adopt the ``best" (i.e., spectroscopic if present, otherwise photometric) redshift presented in C16 (of which 1,981 are spectroscopic and 1,307 are photometric). We adopt the C16 redshift to ensure that we can use their derived X-ray luminosities in our analysis. Next, we select galaxies in the redshift range $0.05 \leq z < 2.5$, leaving 783,028 sources. This redshift range includes the vast majority of detections in the \textit{Herschel PEP} survey, as the detection fraction drops off considerably at redshifts greater than $z=2.5$ (see Figure 12 from \citealt{Lutz11}). Importantly, however, this redshift range spans the epoch during which the majority of SMBH and galaxy growth took place \citep{Aird10, Delvecchio14}.
  \item We then derive stellar masses for all our remaining sources by fitting their SED using Code Investigating GAlaxy Emission \citep[CIGALE,][]{Noll09, Serra11}. Our choice of using CIGALE for the X-ray detections is based on its ability to include the presence of emission from an AGN in its SED fitting routine. To avoid introducing a bias we also recalculated stellar masses for all the remaining sources rather than use the stellar mass presented in L16 (with the AGN component switched off). We provide more details in calculating stellar masses in Section \ref{mass}. We then select only those sources with $\log_{10} (M_*/M_\odot) \geq 10.5$ to ensure the sample is mass-complete across our entire redshift range. This leaves us a sample containing 58,241 galaxies.
  \item Next, we obtain 2-10 keV luminosities (or upper limits thereof) for the remaining sources. Where the source is present in C16, we adopt the luminosity (or upper limit) given in that catalogue. If the source is not detected we calculate a 2-10 keV luminosity upper limit using the sensitivity maps of the \textit{Chandra}-legacy survey (Civano, priv. comm.). How these upper limits are calculated is fully explained in Section~\ref{uplim}. Any of the 58,241 sources in our sample that are not covered in the sensitivity map are deemed to have insufficient X-ray data and thus removed, leaving a sample of 40,418 (of which 2,763 have a measured X-ray luminosity).
  \item SFRs in this sample are calculated by fitting SED models on IR to radio photometry taken from \cite{Jin18}. The photometry catalogue is produced by a ``super-deblending" technique \citep{Liu18}, including de-confused photometry at MIPS/24$\mu m$, \textit{Herschel}, SCUBA2, AzTEC and MAMBO wavelengths, supplemented by NIR Ks, IRAC (L16) and radio data \citep{Smolcic17, Daddi17}. We used the same SED fitting algorithms described in \cite{Liu18}, included AGN models of \cite{Mullaney11} and the spectroscopic redshifts of C16 to ensure redshift consistency. We then classified the sources according to the ``starburstiness" quantity as described in \cite{Schreiber15}. This calculation is further explained in Section~\ref{sfrsect}. Sources without radio or MIPS/24$\mu m$ data are omitted as a radio or MIPS/24$\mu m$ detection is required for the deblending routine. The non-detection at these wavelengths could indicate a lower SFR and such sources are, therefore more likely to be classified as non-starburst. Whilst we could include these sources in our analysis under this assumption, our non-starburst sample is already the larger of the two samples in all of our redshift bins sized and thus does not warrant the introduction of such an assumption. After removing those galaxies without radio or MIPS/$24\mu m$ detections, our final sample size is 26,419.
  \item Finally, in order to investigate any redshift evolution in our $sL_{\rm X}$ distributions we subset our sample into three redshift bins: $0.05 \leq z < 0.5$, $0.5 \leq z < 1.5$ and $1.5 \leq z < 2.5$. The number of detected and upper limits for each redshift bin can be seen in Table~\ref{datatable}. In addition, Figure~\ref{hist_ulfrac} shows the detected $sL_{\rm X}$ distribution for both the starburst and non-starburst samples for each redshift bin and the cumulative upper limit fraction.
\end{enumerate}

\begin{table*}
  \centering
  \label{datatable}
  \begin{tabular}{|l|c|c|c|c|c|c|}
  \hline
                                      & \multicolumn{2}{c|}{$0.05 \leq z < 0.5$} & \multicolumn{2}{c|}{$0.5 \leq z < 1.5$} & \multicolumn{2}{c|}{$1.5 \leq z < 2.5$} \\ \hline
                                      & Detected          & Upper Limit          & Detected          & Upper Limit         & Detected          & Upper Limit         \\ \hline
  \multicolumn{1}{|c|}{Starburst}     & 10                & 97                  & 54                & 516                 & 31               & 227                   \\ \hline
  \multicolumn{1}{|c|}{Non-starburst} & 90                & 1868                 & 780               & 14299               & 461               & 7986                \\ \hline
  \end{tabular}
  \caption{The complete sample sizes for our study, split by redshift bin, starburst classification and whether the sources are X-ray detected or an upper limit on X-ray luminosity had been calculated.}

\end{table*}

\subsection{Stellar masses} \label{mass}

\begin{figure}
  \includegraphics[width=\columnwidth]{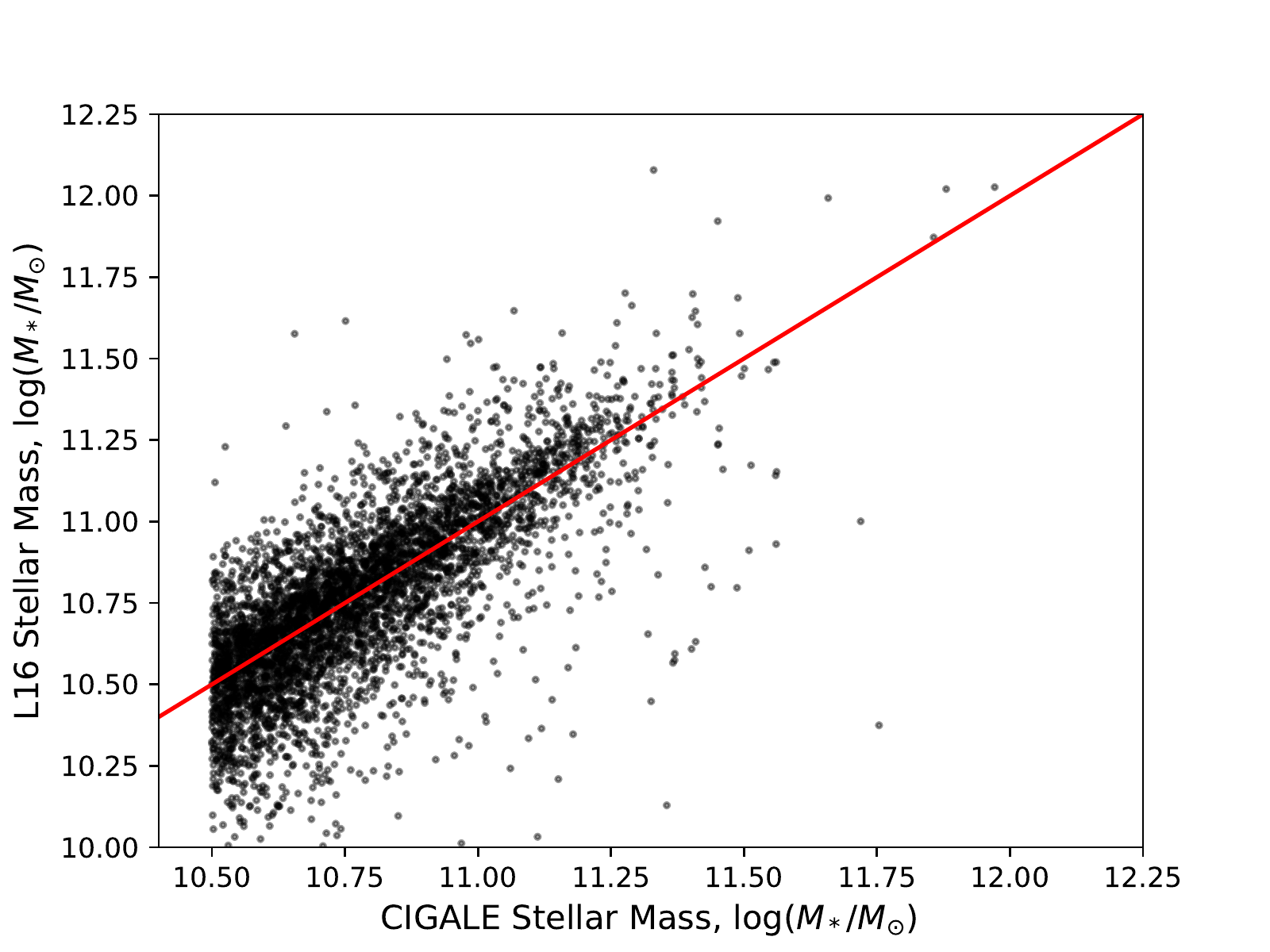}
  \caption{Our CIGALE calculated stellar masses plotted against the mass presented in L16 for 4750 randomly chosen, non X-ray detected sources.}
  \label{massesL16}
\end{figure}

Since we focus on investigating the \textit{specific} X-ray luminosity distribution, we need to calculate accurate stellar masses for all our sources. Host galaxy stellar masses can, however, be difficult to derive accurately in the presence of an AGN \citep{Stern05,Donley12} because of obscuring dust and AGN contamination in the optical-to-near infrared range (i.e., the section of the EM spectrum from which stellar masses are usually derived). Inaccuracies can therefore be introduced if the contribution from the AGN or obscuration due to dust is not accounted-for.

As we need to account for any AGN contamination, rather than adopting the stellar masses presented in L16 we instead use the SED fitting routine CIGALE \citep{Noll09,Serra11,Roehlly14} to independently derive host galaxy stellar masses. CIGALE has the ability to account for AGN contribution by including in the fits the models presented by \cite{Fritz06}, which help to disentangle AGN emission from the host galaxy stellar population. \cite{Ciesla15} studied the ability of CIGALE to reproduce the stellar masses of a mock sample of galaxies and reported that, in the presence of an AGN, the predicted stellar masses were in reasonable agreement with the input. More specifically, the three leftmost plots of Figure~11 in \cite{Ciesla15} highlight the performance of CIGALE for varying quantities of photometric data. Generally, CIGALE performed very well in terms of measuring stellar masses (within 40\% of the input, with no systematic offset) when given photometric data from across the spectrum. We used CIGALE to derive stellar masses for all our sources irrespective of whether they were previously identified (in the X-rays) as hosting AGN, so as to mitigate a calculation bias. 

L16 report photometric data ranging from the far-UV through to the far-IR and so we are confident that we have sufficient data to determine the stellar masses for the sources in our sample. The range of possible parameter values that we used for the CIGALE run are shown in Table \ref{tabcigpar}. These values are chosen as they were found to be the most successful for reproducing stellar masses in \cite{Ciesla15} and are the same values chosen by \cite{Bernhard16}, who highlighted a strong correlation between the masses calculated by CIGALE and those in L16. Figure~\ref{massesL16} shows the good correlation between our CIGALE masses and those in L16 for 4750 randomly-chosen X-ray undetected sources. When computing stellar masses, we switched off the AGN component for sources showing no X-ray detection. Our primary motivation for this is to minimise the number of free parameters in the model. Of course, this means that those AGNs that a not detected in X-rays due to high levels of absorption are modelled as non-AGN galaxies for mass determination, which we acknowledge may introduce systematic errors. We note, however, that since the AGN component is switched off for both the starburst and non-starburst samples, any systematic uncertainties should be broadly consistent between the samples and so we do not anticipate this to have a significant effect on the comparative results. We calculated stellar masses for all sources (within our redshift range) and then chose only those sources with $\log_{10} (M*/M_{\odot}) \geq 10.5$, so that our sample was mass-complete. This leaves us with a sample of 58,241 galaxies. It is also worth noting that \cite{Ciesla15} reported that CIGALE calculated more accurate stellar masses for higher mass galaxies hosting an AGN, where the relative AGN contribution is less significant than at smaller masses (e.g.,$\log_{10} (M*/M_{\odot}) \leq 9.5$), which bolsters the reliability of our derived stellar masses. 

\subsection{X-ray luminosity upper limits} \label{uplim}

If we were to include only X-ray detected sources when measuring our $sL_{\rm X}$ distribution we would be introducing a significant selection bias in to our analysis. It is therefore vital that we include galaxies for which we do not have an X-ray detection by calculating upper limits on their specific X-ray luminosity which we can then include in our maximum-likelihood analysis (see Section \ref{lhoodsect}).

To calculate upper limits on the X-ray luminosities of our sources, we use the 2-10 keV sensitivity map of the \textit{Chandra}-legacy survey (F. Civano, priv. comm.). This provides $3\sigma$ flux upper limits across the whole X-ray coverage of the survey. As such, to obtain flux limits for our non-X-ray detected galaxies we simply extract the flux limit at the position of that galaxy. This corresponds to an \textit{observed} flux limit, whereas for our analysis, we require an intrinsic flux limit that attempts to account for any obscuration due to gas and dust. For detected sources we can use the hard (2-10 keV) to soft (0.5-2 keV) flux ratio to estimate the level of obscuration. This cannot, however, be done for undetected sources so for those we assume an average flux ratio calculated from the detected sources of $Q=1.13$. We acknowledge the possibility that the undetected sources may have a higher level of obscuration than detected sources. However, the distribution of hard to soft flux ratios (for detected sources) is positively skewed. Therefore, the mean is shifted to higher levels of obscuration when compared to the median (~0.74) or mode (~0.53) meaning that the mean value we assume is conservative. In addition, we note there was no significant effect on our results when adopting an even higher obscuration level (e.g, $Q=2$). We then use the following equation to obtain an upper limit on the intrinsic flux based on the upper limit on the observed flux (see \citealt{Bernhard16}):

\begin{equation}
  \log_{10}\left(\frac{F_{\text{I}}}{F_{\text{O}}}\right) = \sum_{i=0}^2 a_i \log_{10}(Q)^i + b_i z^i,
\end{equation}
where $F_{\text{I}}$ is the intrinsic flux, $F_{\text{O}}$ is the observed flux (i.e., the flux limit) and $Q$ is the average flux ratio from the detected sources, i.e., $Q=1.13$. \cite{Bernhard16} found the best fitting values for the coefficients to be $(a_0,\ a_1,\ a_2,\ b_1,\ b_2,\ b_3) = (0.23,\ 0.61,\ 0.041,\ 0.01,\ -0.11,\ -0.02)$, and we adopt these values. After calculating an upper limit on $F_{\text{I}}$, we then use our adopted redshifts to calculate an upper limit on 2-10 keV luminosities, adopting a conversion of

\begin{equation}
	L_x = F_{\text{I}} 4 \pi D^2 (1+z)^{2-\Gamma},
\end{equation}
where $\Gamma = 1.8$ is the assumed averaged intrinsic photon index \citep{Burlon11}.

There are 17,823 galaxies that have insufficient X-ray coverage to calculate a meaningful X-ray upper limit. These are removed from the 58,241 that make up our mass-complete sample leaving 40,418 galaxies, of which 2,763 have a detected X-ray luminosity (the rest have upper limits on X-ray luminosity).  

\subsection{Calculating Starburstiness}\label{sfrsect}

\begin{figure}
  \includegraphics[width=\columnwidth]{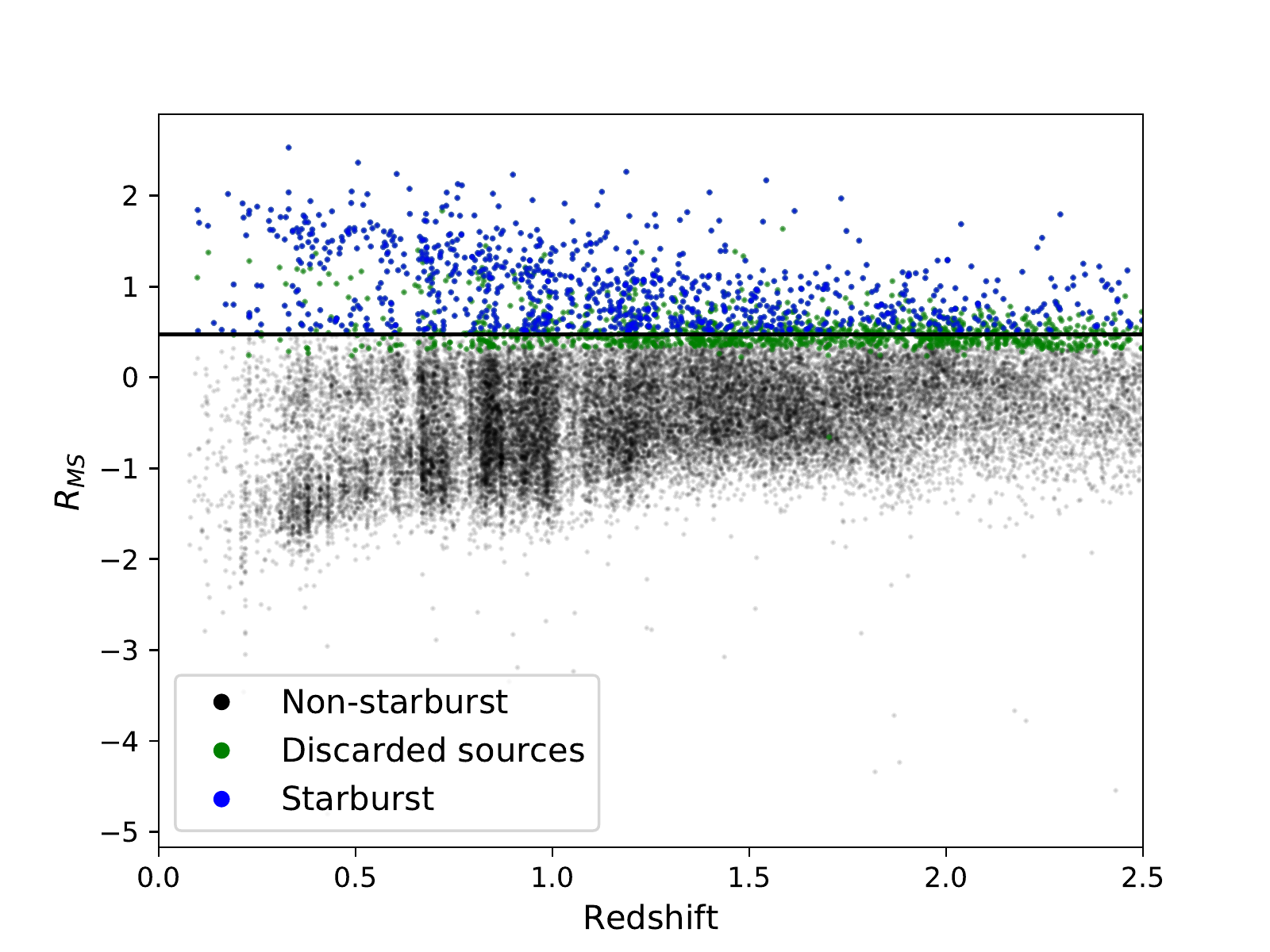}
  \caption{The distribution of specific star formation rate to main sequence (at equivalent mass and redshift) ratio (i.e., $R_{MS}$) as a function of redshift. Sources highlighted in blue are those selected as starburst. Sources in green have been discarded as their uncertainty on SFR estimate could introduce ambiguity into our classification.}
  \label{rms}
\end{figure}

Before we can derive the specific X-ray luminosity distributions we need to divide our sample based on their star-forming properties. In order to do this, we use the catalogue provided by \cite{Jin18}, which provides FIR-based SFRs for the COSMOS field.  \cite{Jin18} adopt a similar deblending routine as that presented in \cite{Liu18}. We use a positional match to identify counterparts in the SFR catalogue to the 40,418 galaxies for our mass-complete sample of galaxies. Since \cite{Jin18} use mostly K-band positions as priors for their deblending we use a small matching radius of 1" to identify counterparts to that catalogue. From these SFRs we calculate specific SFRs (sSFRs) by dividing by the stellar mass of the galaxies. We then classify any galaxy with a sSFR higher than three times that of the main sequence at that redshift as being a starburst galaxy (i.e. $\frac{\rm{sSFR}}{\rm{sSFR}_{\rm{MS}}} \equiv R_{MS} > 3$, see \citealt{Schreiber15}). Since \cite{Jin18} provide uncertainties on SFRs, we choose to discard any sources with ambiguous starburst status (i.e., those galaxies whose SFR error bars span the starburst divide). This prevents the unnecessary introduction of uncertainty. To accurately include information from those sources with ambiguous status a Bayesian hierarchical model would be required, in addition to an analysis without the limitation of binning on SFRs, i.e., an analysis that considers how the $sL_{\rm X}$ changes as a function of SFR, rather than between two bins. Both of these are reserved for a future work. However, as a check, we tested what would happen should we include those sources with ambiguous status assigned based on their calculated starburstiness and noted that it did not have a significant impact on the results. We chose to omit them to minimise the number of potential misclassifications.  Figure~\ref{rms} shows the $R_{MS}$ distribution for all our sources, with starburst sources highlighted in blue and discarded sources in green.

\section{Constructing a flexible model}\label{model}

This study aims to model the full $sL_{\rm X}$ distributions (i.e., including detected and undetected sources) of starburst and non-starburst galaxies in a range of different redshift bins. This section starts by describing how we construct a model that is able to incorporate information from undetected sources, whilst retaining the flexibility required to model the different functional forms the $sL_{\rm X}$ distribution may take. After describing the model, we also derive the likelihood function, from which we can infer the $sL_{\rm X}$ distributions by considering the maximum likelihood estimates of the parameters.

\subsection{Model Selection}\label{selection}

Constraining the precise form of the $sL_{\rm X}$ distribution (or its Eddington ratio equivalent) has been the focus of a number of recent studies \cite[e.g.,][]{Aird17,Bernhard18,Aird18a}. These works have suggested a number of different functional forms for the distribution. Currently, the three most popular functional forms are: a power-law with exponential cut-off \cite[similar to a Schechter function, e.g.,][]{Hopkins09,Aird12,Bongiorno12,Hickox14,Bernhard16,Bernhard18, Wang17, Lanzuisi17, Georgakakis17}, a log-normal distribution \cite[e.g.,][]{Kauffmann09} or a so called ``light-bulb'' model \cite[i.e., accretion is either on or off, e.g.,][]{Conroy13}. The difference in the observed shape of the distribution has recently, however, been attributed to selection effects with \cite{Jones16} suggesting that after correcting for such effects a broad distribution is a good representation for $sL_{\rm X}$ distribution of the AGN population. In this work, we also find that our samples are best modelled by a power-law with exponential cutoff. However, we develop and use a flexible probability distribution that retains the ability to recover both a power-law distribution and, if necessary, a log-normal-like distribution (see Figure~\ref{combgamex}).

In addition to the flexible nature of our model there are a number other criteria that would be desirable for a purpose-built probability distribution. Firstly, we must have a strict probability distribution (i.e., integrates to 1), which enables us to include information from upper limits using the likelihood function (see \ref{lhood} for details). Secondly, for a power-law slope distribution, it is desirable to be able to control the power-law index, and the position of the low and high end exponential cut-offs. In the following subsections, we will describe how our model was built and how we included upper limits into this model.

\subsection{Model construction}\label{lhood}

\begin{figure}
  \includegraphics[width=\columnwidth]{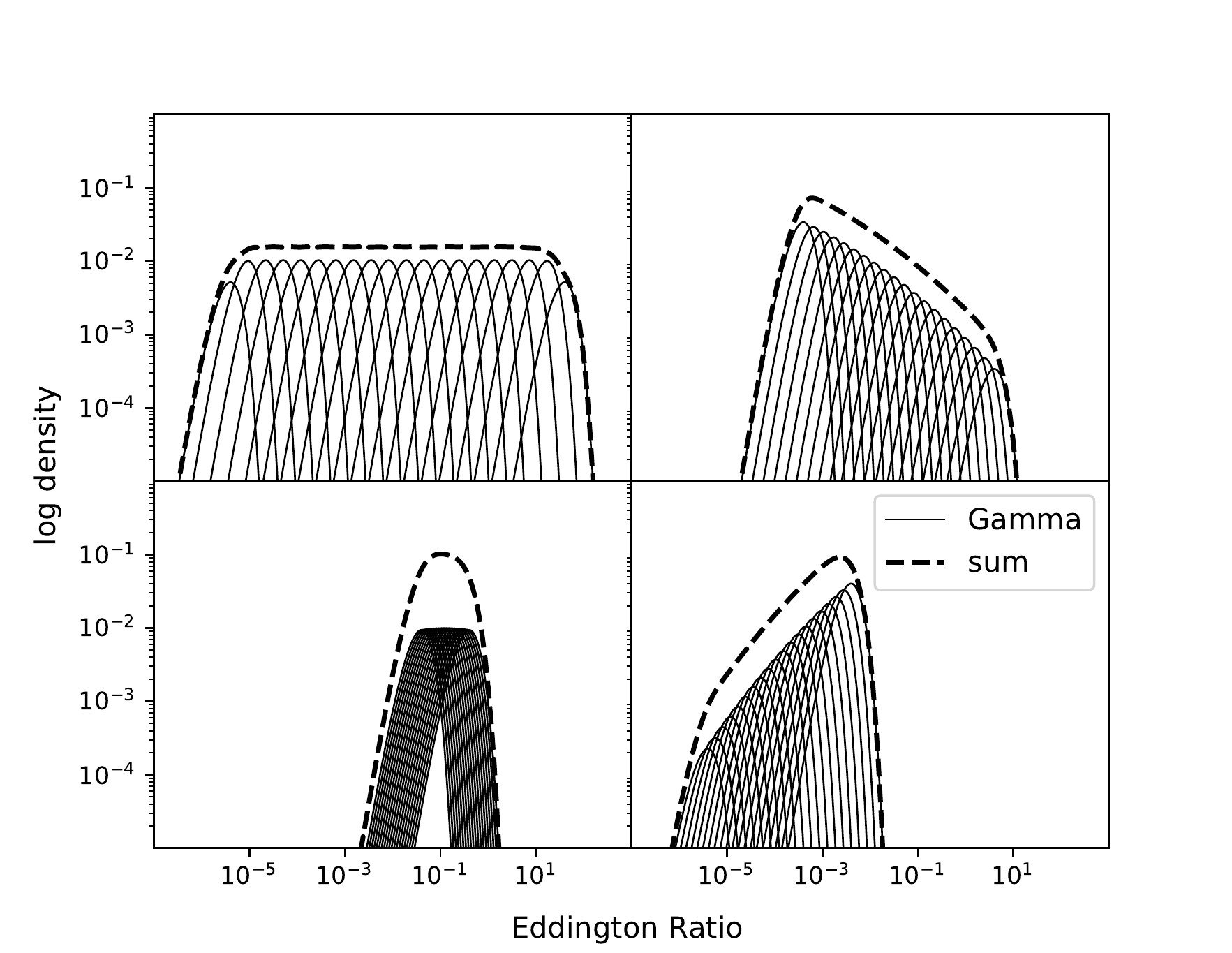}
  \caption{Examples of our model built by the summation of 20 independent gamma distributions (40 are used in the actual model for better accuracy). The parameters are as followed: The shape of each gamma distribution is fixed at $\alpha = 3$. \emph{Top left:} $\gamma = 0$, $\log(\beta_{\text{min}}) = -4$ and $\log(\beta_{\text{max}}) = 1$, \emph{Top right:} $\gamma = -1$, $\log(\beta_{\text{min}}) = -5$ and $\log(\beta_{\text{max}}) = 0$, \emph{Bottom left:} $\gamma = 1$, $\log(\beta_{\text{min}}) = -6$ and $\log(\beta_{\text{max}}) = -1$, \emph{Bottom right:} $\gamma = 0.1$, $\log(\beta_{\text{min}}) = -3$ and $\log(\beta_{\text{max}}) = 1$.}
  \label{combgamex}
\end{figure}

Following \cite{Aird17}, we choose to model our specific X-ray luminosity distributions as a sum of 40 unique Gamma distributions where a single Gamma distribution is described by the following equation:

\begin{equation}
  Ga(X|\alpha,\beta) = \frac{\beta^\alpha}{\Gamma(\alpha)} x^{\alpha-1} e^{-\beta x},
\end{equation}where $\alpha, \beta$ control the position and shape of the distribution and $\Gamma(\alpha)$ is a normalising constant. The mode of the Gamma distribution is given by $\frac{\alpha - 1}{\beta}$. If $\alpha$ is fixed, the mode can be controlled by $\beta$. As such, a set of $\beta$ values can be used to construct a series of equidistant Gamma distributions. If we then take the sum of these Gamma distributions, we recover a flat power-law distribution with lower and upper cut-offs, as seen in the upper-left plot of Figure~\ref{combgamex}. In particular, the minimum value of $\beta$  controls the position of the left-most gamma distribution and the maximum value controls the mode of the right-most. Therefore, controlling the smallest and largest values for $\beta$ allows us to control the positions of the turnovers in our model.

With the position of the lower and upper turnovers controlled by $\beta$ the remaining parameter that we wish to control is the power-law slope. The power-law slope is controlled by the normalisation of the individual gamma distributions. Allocating each gamma distribution with parameter $\beta$ a normalisation (i.e., a multiplicative constant) of $\beta^{\gamma}$ produces a power-law distribution with a slope of $\gamma$ (see Figure~\ref{combgamex}). The lower-left plot in Figure~\ref{combgamex} illustrates how, if the minimum and maximum $\beta$ parameters are close, the model has the ability to fit something similar to a log-normal distribution.

The above model provides us with the flexibility to construct a power-law distribution with appropriate turnovers. Importantly, in addition to this flexibility, summing gamma distributions allows us to easily include information from undetected sources by the incorporation of upper limits. To include upper limits in a likelihood function requires integrating the probability distribution. Using defined parametric distributions, such as the gamma distribution, allows the integrals to be quickly and easily calculated, eliminating the computation time and numerical uncertainties associated with the numerical integration that would be required if we assumed a standard power-law with cutoffs.

\subsection{Likelihood Function}\label{lhoodsect}

Now that we have a description for our model, we need to use our data to obtain the most likely parameter values for our model distributions (hereafter, the parameter values are collectively referred to as $\boldsymbol{\theta} = \{\beta_{\text{min}}, \beta_{\text{max}}, \gamma \}$). For a single X-ray detected galaxy, with $sL_X = x$, the likelihood is given by the probability density function (P.D.F.),

\begin{equation}
  f(x|\boldsymbol{\theta}) = \sum_{i=1}^{40} K \beta_i^{\gamma}\beta_i^\alpha x^{\alpha-1}e^{-\beta_ix},
\end{equation}
where $K$ is a global normalisation constant.

For a sample of $n$ X-ray \textit{detected} galaxies the total likelihood can be written as the product of the probability density functions (P.D.F.), i.e.,

\begin{equation}
  L(\mathbf{\theta}|\mathbf{x}) = \prod_{i=1}^n f(X_i|\mathbf{\theta}).
\end{equation}
In our case, however, we have a large number of non-detections for which we have upper limits on their $sL_{\rm X}$. In such cases we must replace the P.D.F., $f(X_i|\mathbf{\theta})$ with the cumulative distribution function (C.D.F.). That is, the P.D.F. must be replaced by its integral evaluated up to the point of the upper limit. Mathematically, given data $\mathbf{x} = \{X_1,...,X_m,X_{m+1},...,X_n\}$ where $\{X_1,...,X_m\}$ are detected sources and $\{X_{m+1},...,X_n\}$ are upper limits, the likelihood function can now be expressed as,

\begin{equation}
  \label{lhoodeq}
  L(\mathbf{\theta}|\mathbf{x}) = \prod_{i=1}^m f(X_i|\mathbf{\theta}) \prod_{i = m+1} ^n \int_{-\infty}^{\text{UL}_i} f(X_i|\mathbf{\theta}).
\end{equation}

Given a set of $sL_{\rm X}$ for each of the sources in our sample it is this likelihood equation that we seek to maximise. To incorporate uncertainties on the detected sources we calculate an error on the X-ray luminosity by calculating the relative error on the flux observed and propogating this through to the relative error on the luminosity (i.e., neglecting uncertainty on photo-z, for example). For each detected source we then replace the absolute detected value with a randomly sampled value from a Gaussian distribution centred at the observed value with the aforementioned percentage uncertainties. We do this during each step of the maximisation process to accurately account for the uncertainties on $sL_{\rm X}$ throughout the analysis.

\subsection{Likelihood maximisation}

In Section~\ref{lhoodsect}, we derived the likelihood function for our parametric distribution. From here, we can determine which parameter values maximise the likelihood function by using the Markov-chain Monte Carlo Python package \textit{emcee} \citep{ForemanMackey13}. MCMC is required as the likelihood function is too complicated to maximise analytically.

We use MCMC methods to calculate posterior distributions of the parameters of our model, for each redshift bin and both the starburst and non-starburst sample. Our chains each have 200 walkers, each of which are run for 5000 steps (re-sampling the detected values from their uncertainty distributions), with the first 1000 removed for burn-in. This results in a posterior sample of size 800,000 for each parameter. We then choose to thin this sample by selecting every 200th value in the sample. Thinning is used to reduce the sample size to more manageable numbers but also removes the slight dependence between consecutive draws in the chain. On inspection, we noticed the chain converged much more rapidly than the applied burn-in so we are confident we are sampling the posterior parameter space.

\section{Results}\label{results}

We start this section by presenting the output (i.e., the posterior distributions) from the MCMC algorithm. We then discuss the specific parameter results and their potential implications on the $sL_{\rm X}$ distributions for the starburst and non-starburst samples.

\subsection{MCMC output}
\begin{figure}
  \includegraphics[width=\columnwidth]{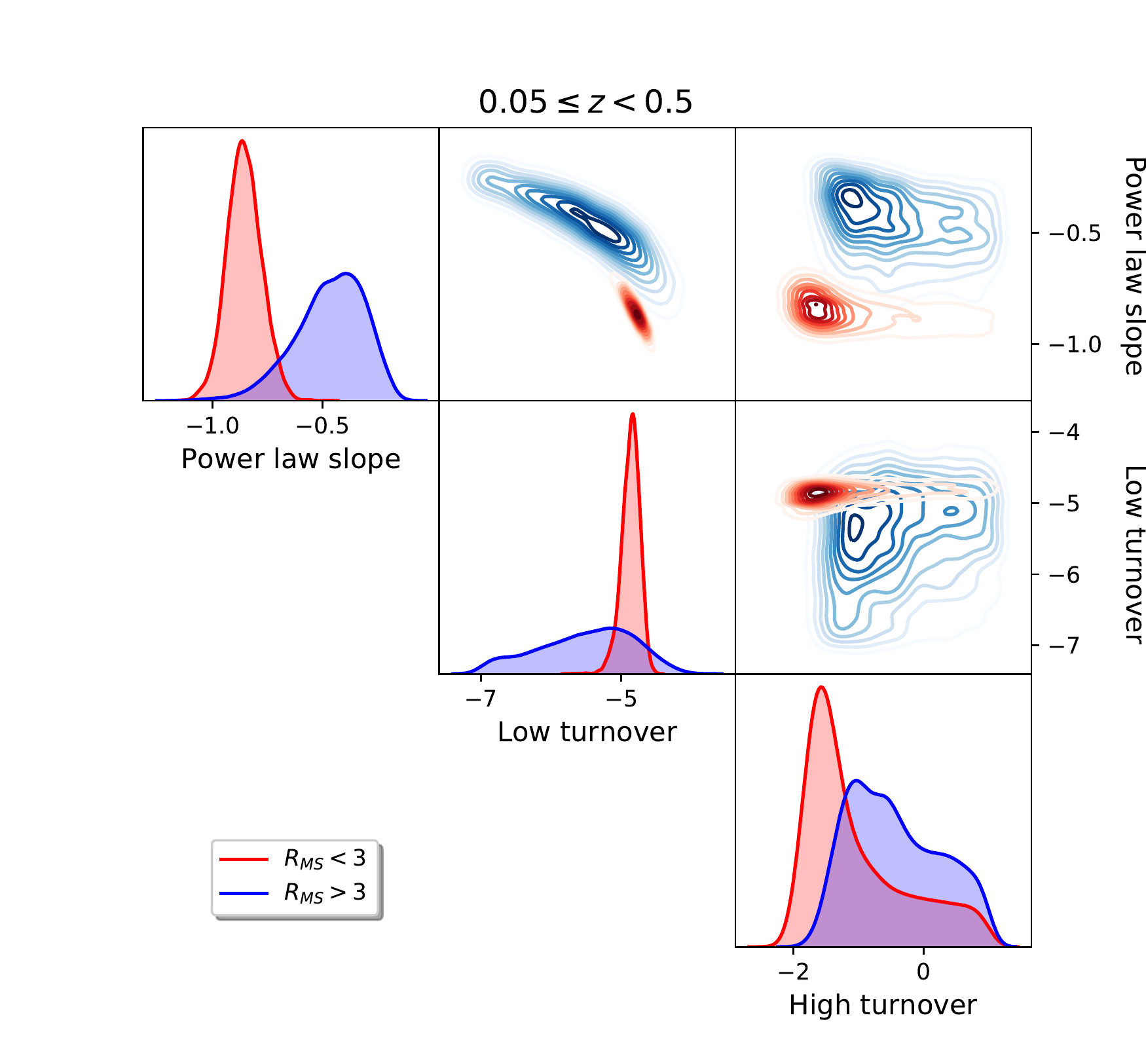}
  \centering
  \caption{The posterior distributions (on diagonal) and the 2-D contour plots, drawn using a kernel density estimation technique for the redshift range $0.05 < z < 0.5$ split between starburst (blue) and non-starburst (red).}
  \label{contour_z05_3}
\end{figure}

\begin{figure}
  \includegraphics[width=\columnwidth]{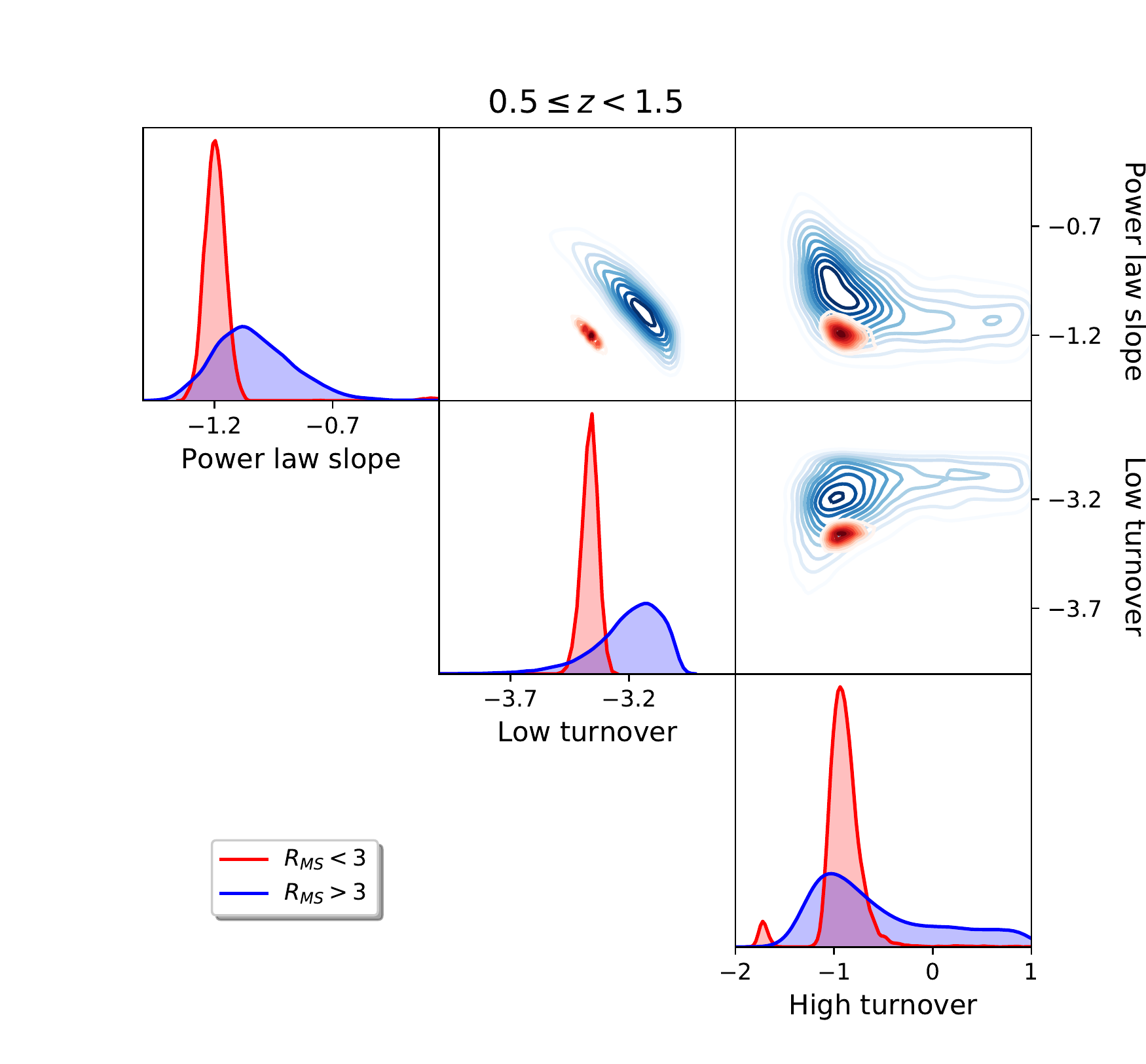}
  \centering
  \caption{Same as Figure \ref{contour_z05_3}, but for the redshift range $0.5 < z < 1.5$.}
  \label{contour_05z15_3}
\end{figure}

\begin{figure}
  \includegraphics[width=\columnwidth]{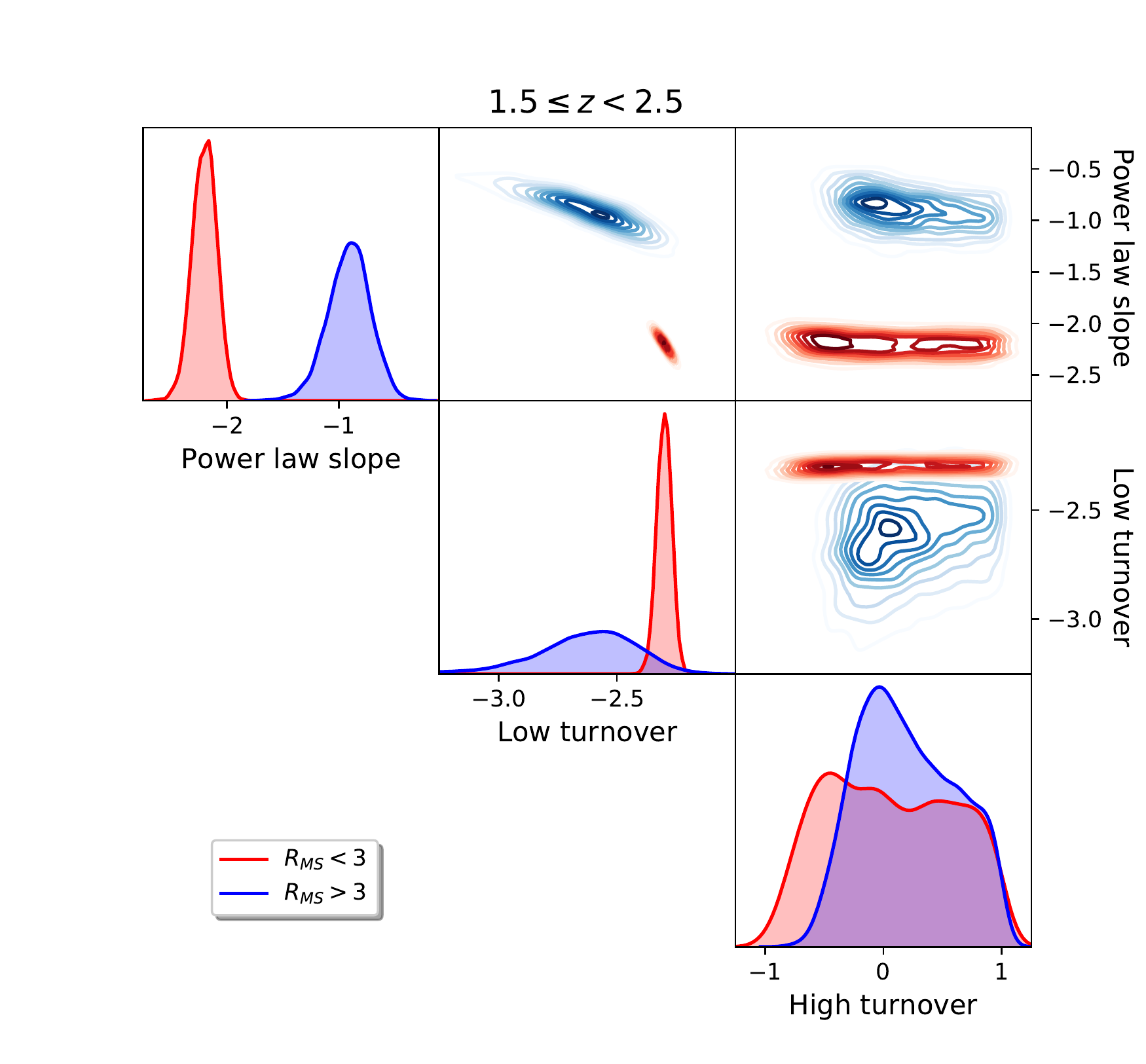}
  \centering
  \caption{Same as Figure \ref{contour_z05_3}, but for the redshift range $1.5 < z < 2.5$.}
  \label{contour_15z_3}
\end{figure}

We present the burned-in, thinned, posterior distributions for the three redshift bins, $0.05 \leq z < 0.5, 0.5 \leq z < 1.5$ and $1.5 \leq z < 2.5$ for both starburst and non-starburst sources in Figures~\ref{contour_z05_3}, ~\ref{contour_05z15_3} and ~\ref{contour_15z_3}, respectively. They show repeated MCMC draws from the posterior distribution of each parameter on the diagonal, as well as the 2D contour plots (shown because of the potential dependence between model parameters) on the off-diagonal, calculated using kernel density estimation (a non-parametric way of estimating a distribution from a histogram using smoothing). In this figure, as well as all further plots, the starburst sample is shown in blue, whereas the non-starburst sample is shown in red. Summary statistics from the posterior samples are shown in Table~\ref{sumstat}.

\begin{table*}
  \centering
  \label{sumstat}
  \begin{tabular}{|c|c|c|}
  \hline
                  & Starburst                    & Non-starburst               \\ \hline
  \multicolumn{3}{|c|}{$0.05 \leq z < 0.5$}                                    \\ \hline
  power law slope & -0.406 (-0.571, -0.275) & -0.857 (-0.944, -0.791)  \\ \hline
  low turnover    & -5.161 (-6.01, -4.641) &  -4.842 (-4.877, -4.734) \\ \hline
  high turnover   & -1.194 (-1.429, 0.016) & -1.610 (-1.929, -0.808)  \\ \hline
  \multicolumn{3}{|c|}{$0.5 \leq z < 1.5$}                                     \\ \hline
  power law slope & -1.090 (-1.212, -0.900) & -1.203 (-1.248, -1.160) \\ \hline
  low turnover    & -3.138 (-3.257,-3.017) & -3.377 (-3.395, -3.328) \\ \hline
  high turnover   & -1.126 (-1.357, -0.332) & -0.965 (-1.051, -0.799)\\ \hline
  \multicolumn{3}{|c|}{$1.5 \leq \ < 2.5$}                                     \\ \hline
  power law slope & -0.902 (-1.077, -0.711) & -2.178 (-2.301, -2.084) \\ \hline
  low turnover    & -2.518 (-2.781, -2.389) & -2.303 (-2.332, -2.268) \\ \hline
  high turnover   & -0.051 (-0.314, 0.553) & -0.556 (-0.614, 0.608) \\ \hline
  \end{tabular}
  \caption{Modes from the posterior distributions presented in Figures~\ref{contour_z05_3}, ~\ref{contour_05z15_3} and ~\ref{contour_15z_3}. The errors, displayed in brackets, are the 68\% highest posterior density intervals calculated using the \textit{HPDInterval} package in R.}
\end{table*}

\begin{figure*}
  \includegraphics[width=\linewidth]{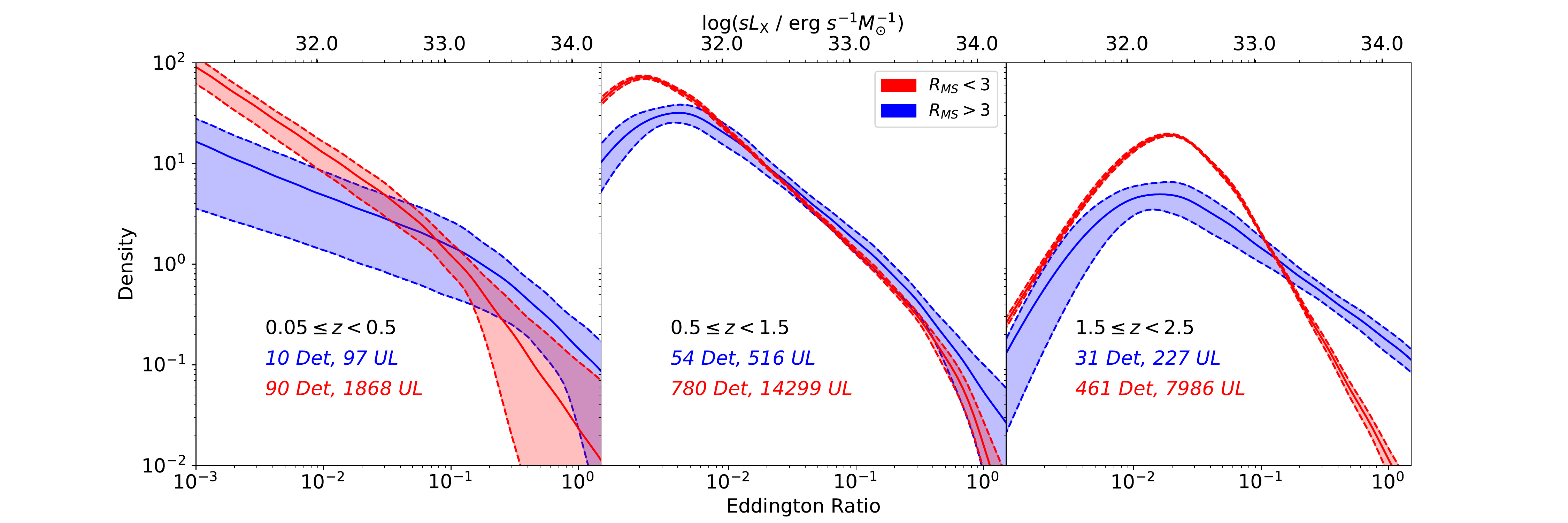}
  \centering
  \caption{The full $sL_{\rm X}$ distributions inferred from our analysis for all three redshift bins. The $1\sigma$ error regions are shown by the shaded region (calculated by finding the 16th and 84th percentile at a fixed value of $sL_{\rm X}$). It should be stressed that these error regions are not errors on the whole distribution, rather on a given value for $sL_{\rm X}$. The starburst sample is shown in blue, while the non-starburst are shown in red. The sample sizes are also shown for reference. }
  \label{alldistplots}
\end{figure*}

By randomly selecting from the posterior parameter values we can construct the range of possible $sL_{\rm X}$ distributions. This is shown in Figure~\ref{alldistplots}, in which we highlight the median $sL_{\rm X}$ distributions including $1\sigma$ error regions, for the three redshifts bins. The errors are calculated by identifying the 16th and 84th percentiles at a given value of $sL_{\rm X}$ for all the sampled parameter values. In the following subsections we discuss, in more detail, the differences between the parameter values for the two starburst samples and as a function of redshift. As is good statistical practice, the posterior distributions displayed in Figure~\ref{alldistplots} are only displayed between the range of the minimum and maximum values of detections.

\subsection{Power law slope} \label{subsec:plslope}

The power law slope parameter controls the gradient of the model between the low and high exponential turnovers. The steepness of this slope could be indicative of the proportion of very luminous sources in the sample, because the slope largely controls the ratio of higher to lower $sL_{\rm X}$ sources (i.e., above and below the midpoint, respectively). From the posterior distributions presented in the upper-left plots of Figures~\ref{contour_z05_3}, \ref{contour_05z15_3} and ~\ref{contour_15z_3}, we see consistently that the modes of the power law slope distribution are shifted to less negative values for the starburst samples in all three redshift bins. In the lowest, intermediate and highest redshift bins we can state that the power law slope in starburst galaxies is shallower than in non-starburst galaxies at a significance of 97.7\%, 80.9\% and 98.5\% respectively. This could suggest that the proportion of higher $sL_{\rm X}$ sources is greater in the starburst population than the non-starburst population (as a result of having a higher ratio of high to low $sL_{\rm X}$ sources) and we explore this possibility further in Section~\ref{inf}. The difference in power law slope can also be seen in the full posterior $sL_{\rm X}$ distributions shown in Figure~\ref{alldistplots} with the gradients of the distributions prior to the break displaying a greatest difference in the high redshift bin. 

\subsection{High turnover} \label{subsec:ht}

Whilst the power law slope indicates the ratio of high to low $sL_{\rm X}$ sources (above and below the midpoint), the high turnover controls the maximum possible values of $sL_{\rm X}$ in the model. From the posterior distributions presented in the lower-rightmost plots of Figures~\ref{contour_z05_3}, \ref{contour_05z15_3} and ~\ref{contour_15z_3}, we see that there is significant overlap between the high turnover distributions in both samples across all the redshift bins. We see a shift in the mode of the posterior distributions in our lowest and highest redshift bins. In addition to this, the high turnover posterior distributions are generally broader than those of the power law slope. We suspect that this is a consequence of this extreme end of the model being constrained by extremely luminous, extremely rare AGN and therefore the inferred posterior distribution is poorly constrained. Having said that, in the highest redshift bin, the significant difference in power-law slope and the inability to recover the high turnover accurately enough combines to create an excess of very high $sL_{\rm X}$ sources in the starburst sample, as shown in Figure~\ref{alldistplots}. Therefore, at this high redshift we cannot rule out that SMBHs in starburst galaxies have the ability to accrete at higher maximum thresholds.

\subsection{Parameter evolution with redshift}

\begin{figure*}
  \includegraphics[width=\linewidth]{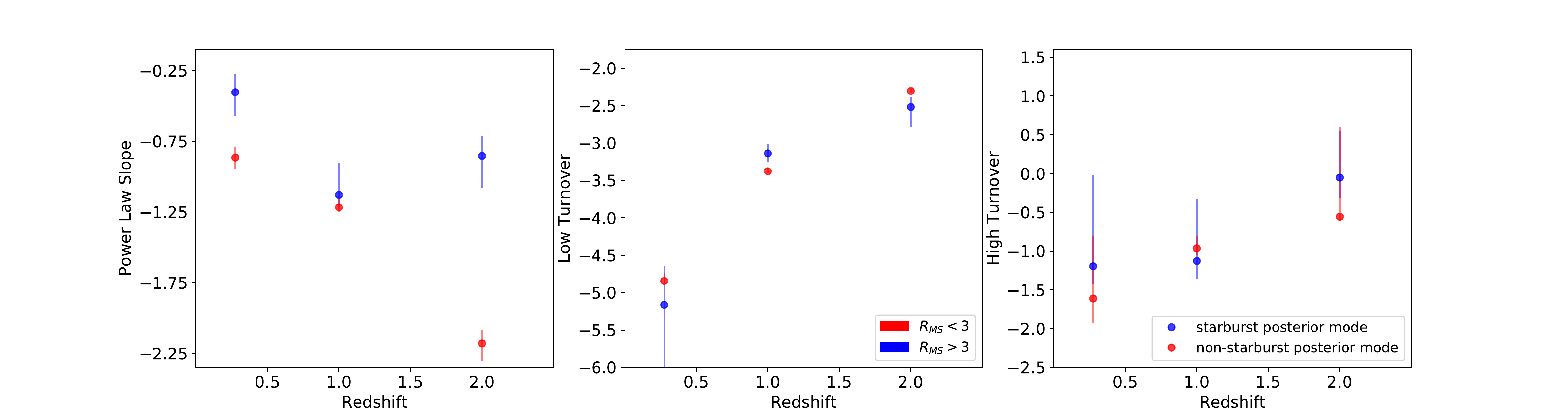}
  \centering
  \caption{Parameter evolution plots for each redshift bin between the staburst (blue) and non-starbursts (red). The posterior mode for each parameter is plotted against the midpoint of the redshift bin it has been inferred from, along with $1\sigma$ uncertainties.}
  \label{paramevolution}
\end{figure*}

As previously mentioned, we subset our sample into three redshift bins to investigate how the various parameters describing our distributions evolve from a redshift of $z\sim2.5$. In Figure~\ref{paramevolution} we show how the mode of the posterior distributions change for each parameter as a function of redshift. Figure~\ref{paramevolution} shows the mode of the posterior distributions for each parameter (power law slope, low turnover and high turnover in the left, middle and right plots, respectively) plotted against the midpoint of the redshift bin it was inferred from. 

The leftmost plot in Figure~\ref{paramevolution} shows how the power law slope has evolved with redshift. This plot suggests that the power law slope os non-starbursts becomes more negative as we go to higher redshifts. As the power law slope may reflect the ratio of higher to lower (i.e., above and below the \textit{mid-point}) $sL_{\rm X}$ sources, the apparent parameter evolution indicates that the proportion of higher $sL_{\rm X}$ sources in the non-starburst galaxy population may have also evolved with redshift. More specifically, as the power law slope has declined out to higher redshifts, the proportion of higher $sL_{\rm X}$ in AGN has also declined. In addition to this, the difference in parameter evolution between the two samples suggests that the proportion of high $sL_{\rm X}$ sources was higher in the starburst population than the non-starburst one, which indicates that a relationship between intense star formation and galaxy evolution is likely to exist and that the evolution of this parameter is more dependent on starburst/non-starburst classification than redshift. However, considering how the low and high turnovers (indicating the range of $sL_{\rm X}$) evolve alongside the power law slope will provide a more complete picture. The low turnover rapidly increases with redshift and whilst the high turnover evolution is poorly constrained (again, due to the rarity of sources at this end of the distribution), it does appear there may be a slight increase with redshift. Should this be the case, it would suggest that while the proportion of higher $sL_{\rm X}$ sources in the population decreases with redshift, the average $sL_{\rm X}$ increases. However, it is worth emphasising that the middle and right plots in Figure~\ref{paramevolution} do suggest that the difference in the low and high turnover between the accretion rate distributions is primarily driven by redshift and not starburstiness, whereas the left plot suggests a greater dependence on starburstiness. We explore the implications of this in Section~\ref{inf}.

\section{Discussion}\label{discussion}

The primary goal of this study is to measure the differences, if any, in the distributions of SMBH accretion rates for starbursting and non-starbursting galaxies. We used the specific X-ray luminosity (i.e., $sL_{\rm X} = L_{\rm X}/M_{\ast}$) as a proxy for Eddington ratio and derived the SFR of our sources from \textit{Herschel} FIR photometry via a deblending and SED fitting routine (see \citealt{Jin18} for details). Our sources were split according to their star-forming properties; if their star formation rate placed them a factor of three above the main sequence (using the prescription of \citealt{Schreiber15}) they were classed as a starburst galaxy, otherwise they were classed as a non-starburst. 

\subsection{Assumptions and analysis limitations}
In order to model the distribution of non-starburst and starburst galaxies as accurately as possible we constructed a flexible parametric model that was able to recover either of the two most popular forms of the $sL_{\rm X}$ distribution reported in the literature (see Section ~\ref{selection}). However, the model is not without limitations and we acknowledge and discuss these further in this section.

Firstly, as with any parametric study, our analysis and interpretation of results are model dependent. A parametric form of the distribution must be assumed (in this case, a power-law with exponential cut-offs or log-Gaussian) in order to account for information from both detections and non-detections. The aim of the study is then to derive the most-likely parameter values for a given model and compare those parameters between samples. From that, we can first pose the question: given our model, do the parameters that describe the underlying distributions differ {\it significantly} for our starburst and non-starburst samples? If so, then the {\it underlying} distributions differ. If they do differ, then we can also ask, given our model, how to they differ? It is important, however, to consider the limitations of our (or any other) model, particlarly when considering the latter question. For example, we acknowledge that our model is incapable of replicating the distribution found in \cite{Aird17}, who found a ``bump" in the distribution at lower $L_{\rm X}$ values ($10^{39} - 10^{41} \text{erg s}^{-1}$ depending on mass and redshift) that they attributed to star formation. As such, any differences in our inferred distributions could be due such a bump that we do not specifically model. However, were we to include a bump at lower $sL_{\rm X}$ values, it would likely cause the inferred power law slope of our starburst sample to flatten further (as upper limits would occupy the bump) which would strengthen the significance of our results.

Secondly, the data in this study contains a large fraction of non-detections. The reason for this is that we intend to infer our results on the entire galaxy population as opposed to only X-ray detected sources, as the latter would produce biased results. However, aside from the appeal of an unbiased sample, the non-detections do contain information about the underlying distribution. The cumulative distribution function (C.D.F.) used in this analysis allows us to incorporate information from the non-detections by fully considering the possible values for them. In Figures~\ref{contour_z05_3}, ~\ref{contour_05z15_3} and ~\ref{contour_15z_3}, one can see the power law slope and the low turnover are correlated. One possible reason for this is that initially, at the high $sL_{\rm X}$ end of the distribution, the power law slope is inferred from the detected sources and the model then computes whether enough upper limits are introduced to maintain this slope. This indicates that our model is sensitive to the fraction of upper limits in the analysis (the low turnover must occur at the point where upper limits are unlikely to be able to maintain the gradient of the most likely power law slope, which is inferred from the detections). As such, it is likely that the low-turnover at low Eddington ratios is a direct consequence of the combination of large numbers of upper limits in the data with our assumed model shape. This further stresses the importance of ensuring that we have a sample representative of the population with a proportionate fraction of non-detections and a justified choice of model.

As with any population study, it is extremely difficult to rule out all possible systematic effects that could influence our final results. We attempt to mitigate the effects of any unknown systematics by (a) treating starburst and non-starburst samples the same in terms converting X-ray fluxes to accretion rates and (b) comparing starbursts to non-starbursts within the same redshift bin and thus minimising the influence of, e.g., flux limits between the samples. Considering point (a) specifically: one could imagine that starburst galaxies have a higher level of absorption due to enhanced amounts nuclear gas introduced by galaxy interactions. If this were the case, then this would work to enhance the differences we see, as correcting for stronger absorption in starbursts would systematically increase the intrinsic $sL_{\rm X}$ we measure, leading to an even greater number of high $sL_{\rm X}$ AGN amongst starbursts.

\begin{figure*}
  \includegraphics[width=\linewidth]{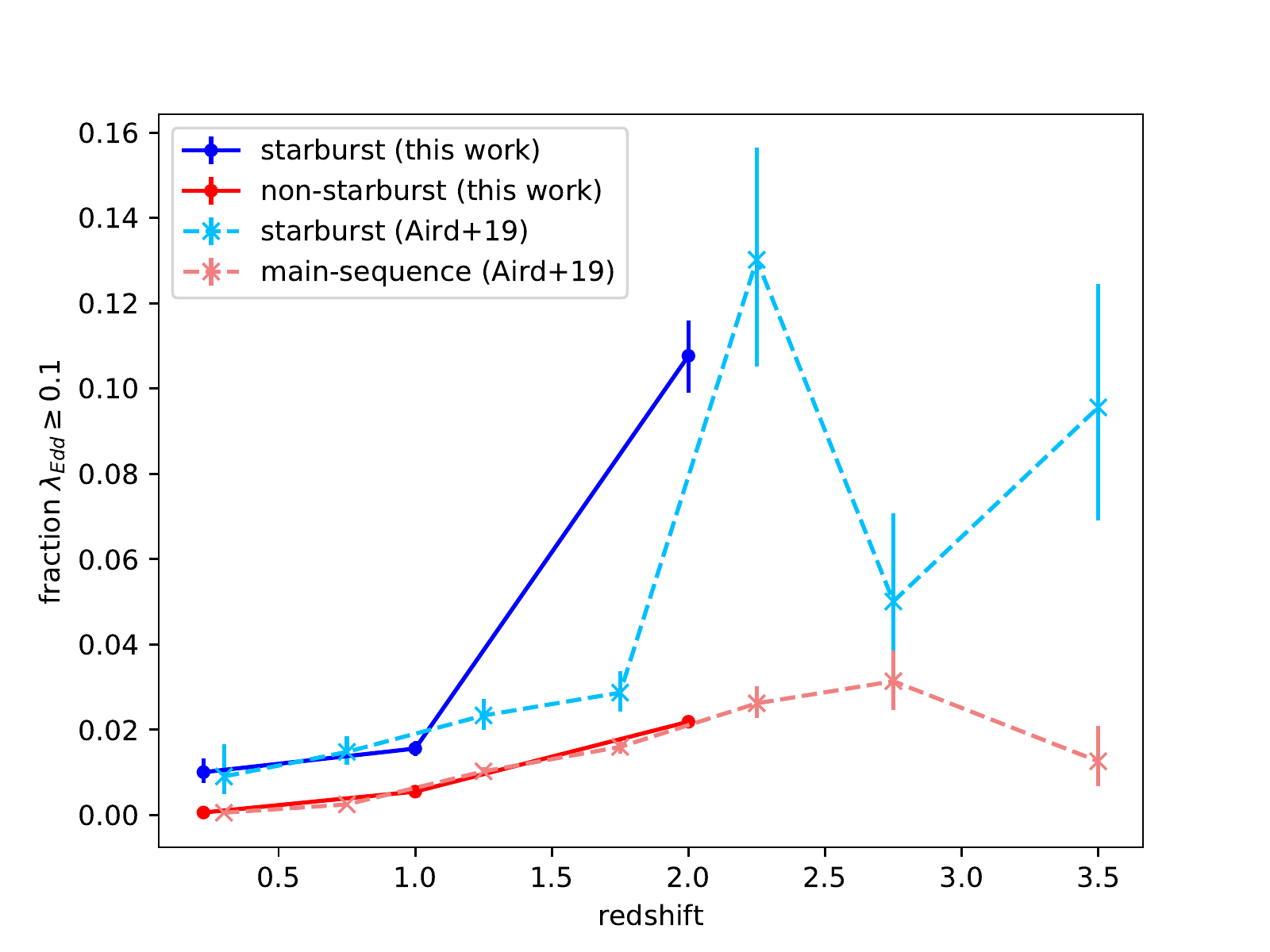}
  \centering
  \caption{Fraction of sources with high accretion rates (i.e., greater than $0.1\lambda_{\text{Edd}}$) as a function of redshift for the starburst and non-starburst samples. Uncertainties are $1\sigma$ and are calculated by selecting the 99.7\% credible interval from the posterior $sL_{\rm X}$ distributions. Overplotted are the starburst and main-sequence fractions from \citet{Aird18b} with $1\sigma$ uncertainties.}
  \label{boxplot}
\end{figure*}

\subsection{Inferring the results} \label{inf}

Figures~\ref{contour_z05_3}, ~\ref{contour_05z15_3} and ~\ref{contour_15z_3} suggest that the parameter with the largest difference between the starburst and non-starburst samples is the power law slope. Given our model, the probability that the accretion rate distribution for starburst galaxies has a less-negative power law slope in the lowest, middle and highest redshift bins are 97.7\%, 80.9\% and 98.5\% respectively. While these differences are not significantly different at the $3\sigma$ level, a difference in the power law slope may indicate that the fraction of higher $sL_{\rm X}$ sources may be different between the starburst and non-starburst samples at a given value of specific X-ray luminosity or Eddington ratio. In order to investigate this further, we calculated the fraction of sources with ``high'' accretion rates (i.e., greater than $0.1\lambda_{\text{Edd}}$) in both the starburst and the non-starburst posterior accretion rate distributions. This is calculated by integrating each of the 4000 posterior $sL_{\rm X}$ distributions for each sample from $0.1\lambda_{\text{Edd}}$ upwards.  These fractions are presented in Figure~\ref{boxplot} and show that the starburst sample has a larger fraction of sources with high accretion rates across all redshift bins. Also included in this plot are the ratios of high to low accretion rate AGN for starbursts and main-sequence galaxies derived from the $sL_{\rm X}$ distribtions of \cite{Aird18b}.\footnote{\cite{Aird18b} used optical to near-infrared SED fits, as opposed to the far-infared data used in this study, to classify galaxies according to their star-forming properties} We note a remarkable consistency between our results and those of that earlier work. 

In order to be able to quantify the difference in these fractions we calculate the probability that a randomly selected posterior $sL_{\rm X}$ distribution from the starburst sample has a higher fraction of high accretion rate sources than a randomly selection posterior distribution from the non-starburst sample. We find that starbursts have a larger fraction of high accretion rate AGN than non-starbursts in 99.6\%, 99.97\%, and >99.99\% of cases in our low, middle, and high redshift bins, respectively. In other words, our inferred distributions suggest one is significantly more likely to identify a high accretion rate AGN in a given starburst compared to a given non-starburst. 

The result that the starburst population has a higher fraction of high $sL_{\rm X}$ is consistent with the findings of \cite{Georgakakis14, Bernhard19, Wang17, Aird17, Aird18a}, who found that the distribution of accretion rates was shifted to lower values in quiescent galaxies compared to star-forming galaxies. By contrast, we also find no strong evidence that the positions of the exponential turnover in the distribution differs between the two populations. Overall, we interpret this in terms of SMBHs in starburst galaxies spending longer at higher accretion rates, but the maximum possible accretion remain broadly the same across the two populations. This could be caused by the SMBH self-regulating at accretion rates close to the Eddington limit (i.e., when the radiation pressure and gravitational forces are in equilibrium). With recent evidence that starburst events are more commonly associated with interactions \citep{Pawlik18, Kauffmann18, Dietrich18} this could be interpreted as further evidence that interactions also enhance the levels of SMBH accretion \citep{Comerford15,Glikman15, Ricci17}.

At face value, our results seem to show no indication that intense radiation produced from an AGN during an accretion phase negatively impacts star-formation \citep{DiMatteo05,Fabian12}. Otherwise, we may have expected to find heightened accretion rates within non-starburst galaxies. However, care must be exercised when considering the stochastic nature of AGN variability, since any impact on the FIR-derived SFR will be delayed by roughly 100 Myr \citep{Kennicutt98}. Indeed taking the complementary approach of measuring the SFR distribution in X-ray luminosity bins, \cite{Scholtz18} demonstrates the need for negative feedback in simulations to reproduce the observed X-ray luminosity-dependent stellar mass specific SFR (sSFR) distributions. This demonstrates that the relationship between AGN feedback and SFR requires multiple complementary analysis methods to provide a complete picture. We therefore stress that the above result should not be interpreted as evidence against AGN activity quenching star formation rate, as any study of this nature fails to adequately account for the time-delay between AGN activity and the shutting-down of star formation.

\section{Conclusions}

In this study we have developed a flexible model in order to infer the specific accretion rate distributions of central SMBHs within starburst and non-starburst galaxies. Our model distribution consists of a power-law curtailed by an upper and lower turnover, and allows us to incorporate information from upper limits, thereby allowing our sample to be more representative of the galaxy population in general. We derived the specific accretion rates from the 2-10~keV X-ray luminosities (or upper limits thereof) and used deblended \textit{Herschel} maps to estimate the star formation rates. A source was classified as starburst if it had a SFR a factor of 3 greater than the main sequence at its redshift.

The main conclusions of this work are as follows:

\begin{enumerate}
 \item{Given our assumed model, we find suggestive (i.e., between 1.8 and 3$\sigma$) evidence that the accretion rate distributions for massive galaxies ($\log10(M_*M_{\odot}) > 10.5$) are dependent on both the star-forming properties of the galaxies and on redshift.}
 \item{More specifically, when modelled as a curtailed power-law, the gradient of the power law slope of the accretion rate distribution is shallower (i.e., less negative) in starburst galaxies, suggesting there is a slightly higher probability of detecting a high $sL_{\rm X}$ (high Eddington ratio) AGN in galaxies that have recently undergone an intense period of star-formation. This suggests that SMBHs in starburst galaxies spend more time at higher accretion rates than their non-starburst counterparts.}
 \item{We find stronger evidence that starbursts and non-starbursts differ in terms of their specific accretion rates when we use our posterior $sL_{\rm X}$ distributions to calculate the fractions of such galaxies with high accretion rates (i.e., greater than $0.1\lambda_{\text{Edd}}$). In doing so, we estimate that the fraction of starbursts hosting high accretion rate AGN is larger than the fraction of non-starbursts at confidence levels of 99.6\%, 99.97\%, and >99.99\% for our low ($0.05\leq z<0.5$), mid ($0.5\leq z<1.5$), and high ($1.5\leq z<2.5$) redshift bins, respectively.}
 \item{Within our uncertainties, we find no evidence that the positions of the high end turnover of the accretion rate distribution differs between starburst and non-starburst galaxies. We interpret this as suggesting that, whilst there are a higher fraction of SMBHs accreting at higher rates in the starburst population, the maximum accretion rates do not differ considerably, particularly in our low and middle redshift bins. This suggests that either the SMBHs are being self-regulated as they approach the Eddington limit or at least some other process is preventing accretion at considerably higher rates.}
\end{enumerate}

\section*{Acknowledgements}

We would like to thank the anonymous referee for their detailed report which improved the quality of the paper. We would like to thank Ivan Delvecchio and James Aird for useful discussions and Francesco Civano for the \textit{Chandra} sensitivity maps. LG is supported through a PhD scholarship granted by the University of Sheffield. EB and JM acknowledge STFC grant R/151397-11-1. LG is also grateful to the Astronomy group at the University of Sheffield for continuous helpful discussions. 




\bibliographystyle{mnras}
\bibliography{ref} 




\appendix
\section{Table of SED fitting parameter values}
\begin{table*}
\centering
\begin{tabular}{|c|l|c|}
\hline
\multicolumn{3}{|c|}{\textbf{Modules}}                                                                                                          \\ \hline
\multicolumn{2}{|c|}{\textbf{Process/Model}}                                    & \textbf{Theoretical Model Chosen}                             \\ \hline
\multicolumn{2}{|c|}{Star Formation History}                                    & Delayed                                                       \\ \hline
\multicolumn{2}{|c|}{Stellar Population Synthesis Model}                        & \cite{Bruzual03}                                     \\ \hline
\multicolumn{2}{|c|}{Dust Attentuation}                                         & \cite{Calzetti01}                                                \\ \hline
\multicolumn{2}{|c|}{Dust Emission}                                             & \cite{Dale14}                                                    \\ \hline
\multicolumn{2}{|c|}{AGN}                                                       & \cite{Fritz06}                                                   \\ \hline
\multicolumn{2}{|c|}{Initial Mass Function}                                     & \cite{Chabrier03}                                                \\ \hline
\multicolumn{3}{|l|}{}                                                                                                                          \\ \hline
\multicolumn{3}{|c|}{\textbf{Parameter Values}}                                                                                                 \\ \hline
\multicolumn{3}{|c|}{\textit{Star Formation History}}                                                                                           \\ \hline
\multicolumn{2}{|c|}{e-folding time of the main stellar population model (Myr)} & 100,1000,3000,10000,1E10                                      \\ \hline
\multicolumn{2}{|c|}{Age of the oldest stars in the galaxy (Myr)}               & 100,1000,2000,3000,4000,5000,6000,7000,8000,9000,10000,11000  \\ \hline
\multicolumn{3}{|c|}{\textit{Stellar Population Synthesis Model}}                                                                               \\ \hline
\multicolumn{2}{|c|}{Metallicity}                                               & 0.02                                                          \\ \hline
\multicolumn{2}{|c|}{Separation Age (Myr)}                                      & 10                                                            \\ \hline
\multicolumn{3}{|c|}{\textit{Dust Attenuation}}                                                                                                 \\ \hline
\multicolumn{2}{|c|}{E(B-V)* for the old population}                            & 0.01,0.05,0.1,0.2,0.3,0.4,0.5,0.6,0.7,0.8,0.9,1.0,1.1,1.2,1.4 \\ \hline
\multicolumn{2}{|c|}{E(B-V)* reduction factor of the old population}            & 0.44                                                          \\ \hline
\multicolumn{2}{|c|}{Central wavelength of the UV bump (nm)}                   & 217.5                                                         \\ \hline
\multicolumn{2}{|c|}{Width (FWHM) of the UV bump(nm)}                         & 35.0                                                          \\ \hline
\multicolumn{2}{|c|}{Slope of dust attenuation power law}                         & 0.0, 0.25, 0.5, 0.75                                                        \\ \hline
\multicolumn{3}{|c|}{\textit{AGN}}                                                                                                              \\ \hline
\multicolumn{2}{|c|}{Alpha Slope}                                               & 1.5,2.5                                                       \\ \hline
\multicolumn{2}{|c|}{Ratio of the maximum to minimum radii of the torus}        & 60                                                            \\ \hline
\multicolumn{2}{|c|}{Tau}                                                       & 1.0,6                                                         \\ \hline
\multicolumn{2}{|c|}{Beta}                                                      & -0.5                                                          \\ \hline
\multicolumn{2}{|c|}{Gamma}                                                     & 0                                                             \\ \hline
\multicolumn{2}{|c|}{Full opening angle of the torus(degrees)}                  & 100                                                           \\ \hline
\multicolumn{2}{|c|}{Angle between equatorial axis and line of sight(degrees)}  & 0.001, 89.990                                                 \\ \hline
\multicolumn{2}{|c|}{Fraction of L\_\{IR\} from AGN}                            & 0,\textbf{0.05,0.1,0.2,0.3,0.4,0.5,0.7,0.9}                  \\ \hline
\end{tabular}
\label{tabcigpar}
\caption{The various different modules used and the possible parameter values input into the CIGALE SED fitting code to derive host galaxy properties. Note CIGALE was run differently for those sources with AGN detections. The extra possible parameter values for the AGN run are shown in bold.}
\end{table*}


\bsp	
\label{lastpage}
\end{document}